\documentclass{singlecol-new}
\usepackage[numbers]{natbib}
\usepackage[utf8]{inputenc}
\usepackage{lipsum}
\usepackage{enumitem}
\usepackage{multicol}
\usepackage{booktabs}
\usepackage[flushleft]{threeparttable}
\usepackage{makecell}
\usepackage{tabularx}
\usepackage{cellspace}
\usepackage{svg}
\newcommand{\sectioname}{Section}
\newcommand{\code}[1]{\texttt{#1}}

\begin{document}
\title{Frameworks for SNNs: a Review of Data Science-oriented Software and an Expansion of SpykeTorch}

\authorA{Davide L. Manna}
\affA{Neuromorphic Sensor signal Processing (NSSP) Lab,\\University of Strathclyde,\\ G1 1XW, Glasgow, UK \\
E-mail: davide.manna@strath.ac.uk}
\authorB{Alex Vicente-Sola}

\affB{Neuromorphic Sensor signal Processing (NSSP) Lab,\\University of Strathclyde,\\ G1 1XW, Glasgow, UK}

\authorC{Paul Kirkland}
\affC{Neuromorphic Sensor signal Processing (NSSP) Lab,\\University of Strathclyde,\\ G1 1XW, Glasgow, UK}

\authorD{Trevor J. Bihl}
\affD{Air Force Research Laboratory (AFRL),\\Wright-Patterson AFB, OH, USA}

\authorE{Gaetano Di Caterina}
\affE{Neuromorphic Sensor signal Processing (NSSP) Lab,\\University of Strathclyde,\\ G1 1XW, Glasgow, UK}

\LRH{Manna et~al.}
\RRH{Frameworks for SNNS}
\begin{abstract}
Developing effective learning systems for Machine Learning (ML) applications in the Neuromorphic (NM) field requires extensive experimentation and simulation. Software frameworks aid and ease this process by providing a set of ready-to-use tools that researchers can leverage. The recent interest in NM technology has seen the development of several new frameworks that do this, and that add up to the panorama of already existing libraries that belonged to neuroscience fields. This work reviews 9 frameworks for the development of Spiking Neural Networks (SNNs) that are specifically oriented towards data science applications. We emphasize the availability of spiking neuron models and learning rules to more easily direct decisions on the most suitable frameworks to carry out different types of research. Furthermore, we present an extension to the SpykeTorch framework that gives users access to a much broader choice of neuron models to embed in SNNs and make the code publicly available.
\end{abstract}
\KEYWORD{frameworks, spiking neural networks, spiking neurons, algorithms, neuromorphic, software, data science, machine learning, unsupervised learning}
\maketitle
\section{Introduction}
The development of Deep Learning (DL) algorithms was greatly eased by the introduction of purposely developed software packages. These packages, or frameworks, usually offer a wide range of software tools that aim to speed up the development of Machine Learning (ML) pipelines as well as make the algorithms available to a larger audience. When referring to conventional DL, i.e. non Neuromorphic (NM), several famous libraries exist, such as TensorFlow (TF) \cite{abadi2016tensorflow}, PyTorch \cite{Paszke2019PyTorchanImperativeStyleHighPerformanceDeepLearningLibrary} or Caffe \cite{Jia2014Caffe}. The field of Neuromorphic engineering has recently seen the emergence of several new software frameworks thanks to the renewed interest in its potential. However, these frameworks are often in an early development stage when compared to their conventional DL counterpart, being limited in the tools they offer, their documentation, and the support from the community. Some more established frameworks also exist, but they are often directed towards particular communities and use cases \cite{Schuman2022OpportunitiesforNeuromorphicComputingAlgorithmsandApplications}, or they are neuroscience-oriented frameworks rather than NM-ML development tools. Furthermore, effective data-science algorithms that can close the gap with other conventional methodologies still need to be developed. Indeed, algorithms employing Spiking Neural Networks (SNNs) are already more energy efficient than conventional Convolutional Neural Networks (CNNs) \cite{GarciaVico2021APreliminaryAnalysisonSoftwareFrameworksfortheDevelopmentofSpikingNeuralNetworks}, however, they are not as effective on ML tasks in terms of accuracy. Hence the importance of having good software frameworks that enable customization, simulation and deployment of SNNs. This requires combining a number of key elements into a pipeline such as learning rules, connectivity patterns, and spiking neurons. Regarding the spiking neurons, emerging NM chips such as Loihi 2 \cite{orchard2021EfficientNeuromorphicSignalProcessingWithLoihi2} allow the use of customized models. It has been shown in the literature that different types of neuron models can solve certain tasks more effectively than other models \cite{Frady2022EfficientNeuromorphicSignalProcessingwithResonatorNeurons, Manna2022SimpleandComplexSpikingNeuronsPerspectivesandAnalysisinaSimpleSTDPScenario}. Therefore it can be beneficial for researchers to use a framework that enables seamless experimentation with different types of neurons.

This work contributes by providing a review of data science-oriented frameworks and highlighting the key features they offer. By restricting our review to this kind of frameworks, we hope to help boosting new research in NM for ML applications. Further to this, we develop an expansion\footnotemark{} of the SpykeTorch \cite{Mozafari_2019_Spyketorch} framework that enables the user to experiment on a wider variety of different spiking neuron models. By doing this, we aim to enlarge the scope of the research in SNNs to different spiking neuron models, and to thus build new algorithms that can leverage the latest advances in the NM hardware.

\footnotetext{Code available at https://www.github.com/daevem/SpykeTorch-Extended}
The rest of this paper is organized as follows. \sectioname\ \ref{sec:related} reports about related works and their differences with the current one. \sectioname\ \ref{sec:frameworks} is the review of the software frameworks for SNNs. \sectioname\ \ref{sec:spiking_neurons} reports the key features of the newly implemented spiking neuron models. \sectioname\ \ref{sec:conclusions} concludes the paper. 

\section{Related Works}\label{sec:related}
When presenting a new software framework, authors often report other similar works and draw comparisons with them \cite{Mozafari_2019_Spyketorch, Hazan2018BindsNETaMachineLearningOrientedSpikingNeuralNetworksLibraryinPython}. In these instances, differences in terms of offered features are highlighted, as well as the advantages of using the newly presented software over the existing ones. Other works specifically focus on reviewing the existing frameworks for the development of SNNs. One example is given by \cite{Qu2022AReviewofBasicSoftwareforBrainInspiredComputing}, where the authors make a subdivision of the software packages into three main groups depending on whether they are NM chips toolchains, SNN simulation frameworks or frameworks that integrate SNNs and DNNs. An other work by García-Vico and Herrera \cite{GarciaVico2021APreliminaryAnalysisonSoftwareFrameworksfortheDevelopmentofSpikingNeuralNetworks} gives an introductory overview of SNNs and then reviews some prominent simulation frameworks. The authors also define a simple classification task and compare accuracy and execution time obtained by using the different frameworks. These previous two works consider frameworks regardless of their research orientation, i.e. they consider both neuroscience-oriented and data science-oriented frameworks. In this work we specifically highlight software packages that are data science-oriented and developed in Python or with a Python interface. Furthermore, we also include in our review other different frameworks and highlight some key features and neuron models that they offer for the development of SNNs. 

\section{Software Frameworks}\label{sec:frameworks}
Many of the software libraries for the development of SNNs are oriented toward the needs of the neuroscience and neurobiology fields \cite{GarciaVico2021APreliminaryAnalysisonSoftwareFrameworksfortheDevelopmentofSpikingNeuralNetworks}. Indeed, because SNNs process inputs and communicate information in a way similar to the human brain, they are particularly suited for simulations of brain areas activations. Nevertheless, the recent emergence of NM engineering as a field for the development of ML algorithms has highlighted the need for suitable frameworks. Consequently, following we will present some of the most prominent software packages to develop data science-oriented SNNs along with their main features, which are also summarized in Table \ref{tab:frameworks}.

\begin{table}[t]
\caption{Key elements of the reviewed frameworks. The ``A-'' stands for adaptive, whereas ``H-'' stand for heterogeneous.}
\label{tab:frameworks}

\resizebox{\textwidth}{!}{
\begin{threeparttable}
\begin{tabular}{cccccccccc}
\hline
\textbf{Framework}                                                              & \textbf{Nengo}                                                               & \textbf{Lava}                                                                                  & \textbf{\begin{tabular}[c]{@{}c@{}}SNN\\ Toolbox\end{tabular}}                      & \textbf{Norse}                                                       & \textbf{PySNN}                                                 & \textbf{snnTorch}                                                                  & \textbf{SpikingJelly}                                                 & \textbf{BindsNet}                                                     & \textbf{SpykeTorch}                                                                    \\ \hline
\textbf{\begin{tabular}[c]{@{}c@{}}Spiking \\ Neurons\end{tabular}}             & \begin{tabular}[c]{@{}c@{}}LIF\\ A-LIF\\ IZ\end{tabular}                     & \begin{tabular}[c]{@{}c@{}}LIF\\ R\&F\tnote{*}\\ A-LIF\tnote{*}\\ A-R\&F\tnote{*}\\ A-IZ\tnote{*}\\ $\Sigma-\Delta$\end{tabular} & I\&F                                                                                & \begin{tabular}[c]{@{}c@{}}LIF\\ AdEx\\ EIF\\ IZ\\ LSNN\end{tabular} & \begin{tabular}[c]{@{}c@{}}I\&F\\ LIF\\ A-LIF\end{tabular}     & \begin{tabular}[c]{@{}c@{}}LIF\\ Recurrent LIF\\ 2nd Order LIF\\ LSNN\end{tabular} & \begin{tabular}[c]{@{}c@{}}I\&F\\ LIF\\ pLIF\\ QIF\\ EIF\end{tabular} & \begin{tabular}[c]{@{}c@{}}I\&F\\ LIF\\ A-LIF\\ IZ\\ SRM\end{tabular} & \begin{tabular}[c]{@{}c@{}}I\&F\\ LIF\tnote{**}\\ QIF\tnote{**}\\ EIF\tnote{**}\\ AdEx\tnote{**}\\ IZ\tnote{**}\\ H-Neurons\tnote{**}\end{tabular} \\ \hline
\textbf{\begin{tabular}[c]{@{}c@{}}Learning \\ Rules\end{tabular}}              & \begin{tabular}[c]{@{}c@{}}Oja\\ BCM\\ BP\end{tabular}                       & \begin{tabular}[c]{@{}c@{}}SLAYER\\ STDP\\ 3-Factor  \end{tabular}& Pre-trained & \begin{tabular}[c]{@{}c@{}}SuperSpike\\ STDP\end{tabular}            & \begin{tabular}[c]{@{}c@{}}STDP\\ MSTDP\\ MSTDPET\end{tabular} & \begin{tabular}[c]{@{}c@{}}BPTT\\ RTRL\end{tabular}                                & BP                                                                    & \begin{tabular}[c]{@{}c@{}}STDP\\ Hebbian\\ MSTDPET\end{tabular}      & \begin{tabular}[c]{@{}c@{}}STDP\\ R-STDP\end{tabular}                                  \\ \hline
\textbf{\begin{tabular}[c]{@{}c@{}}Conversion\\ from\end{tabular}}              & TF/Keras                                                                     & PyTorch                                                                                        & \begin{tabular}[c]{@{}c@{}}TF/Keras\\ PyTorch\\ Caffe\\ Lasagne\end{tabular}        & -                                                                    & -                                                              & -                                                                                  & PyTorch                                                               & PyTorch                                                               & -                                                                                      \\ \hline
\textbf{\begin{tabular}[c]{@{}c@{}}Destination\\ Backend/Platform\end{tabular}} & \begin{tabular}[c]{@{}c@{}}Loihi\\ FPGA\\ SpiNNaker\\ MPI\\ CPU/GPU\end{tabular} & \begin{tabular}[c]{@{}c@{}}Loihi\\ CPU/GPU\end{tabular}                                            & \begin{tabular}[c]{@{}c@{}}SpiNNaker\\ Loihi\\ pyNN\\ Brian2\\ MegaSim\end{tabular} & CPU/GPU                                                                  & CPU/GPU                                                            & CPU/GPU                                                                                & CPU/GPU                                                                   & CPU/GPU                                                                   & CPU/GPU                                                                                    \\ \hline
\end{tabular}
\begin{tablenotes}
\item[*] Only available in Lava-DL.
\item[**] Added in this work.
\end{tablenotes}
\end{threeparttable}
}
\end{table}

\subsection{Nengo}\label{sec:nengo}
Nengo \cite{Bekolay2014NengoaPythonToolforBuildingLargeScaleFunctionalBrainModels} is a Python package for building and deploying neural networks. It is composed of several sub-packages to be used in case of  different needs and destination platforms. For instance, NengoDL is to be used when aiming to convert a neural network built using TF/Keras into its Nengo spiking version. NengoLoihi, instead, allows to deploy neural networks natively built in the Nengo Core package onto Loihi chips. Other packages are NengoFPGA, NengoSpiNNaker, NengoOCL and NengoMPI.
Nengo builds on top of a theoretical framework called the Neural Engineering Framework (NEF) \cite{Stewart2012ATechnicalOverviewoftheNeuralEngineeringFramework}. Computations are hence based on the three principles of the NEF, namely neural representation, transformation, and neural dynamics. Neurons in Nengo are organized in Ensembles and different types of neuron models are available, among which the Leaky Integrate-and-Fire (LIF) \cite{Lapicque1907RecherchesQuantitativesSurLExcitationElectriqueDesNerfsTraiteeCommeUnePolarization} and Izhikevich's (IZ)\cite{Izhikevich_2003} models.
Connections between ensembles are designed to allow a transformation of the information from one ensemble to an other.
Training in Nengo is possible with the Oja \cite{oja1982simplified}, BCM \cite{bienenstock1982BCM} and backpropagation (BP) learning rules.
Using Nengo as a tool for the development of SNNs has the main advantage of having the possibility to target a wide variety of backends and to convert conventional DNNs into a spiking equivalent \cite{GarciaVico2021APreliminaryAnalysisonSoftwareFrameworksfortheDevelopmentofSpikingNeuralNetworks}. Nengo also allows for a certain degree of customization of the components; however, it remains very oriented towards the NEF structure.

\subsection{SNN Toolbox}
SNN Toolbox \cite{Rueckauer2017ConversionofContinuousValuedDeepNetworkstoEfficientEventDrivenNetworksforImageClassification} provides a set of tools to perform automated conversion from conventional Artificial Neural Network (ANN) models into SNNs. Conversion is possible from three different DL frameworks, namely TF/Keras, PyTorch, Caffe and Lasagne \cite{lasagne}. The framework supports conversion to models for PyNN \cite{Davison2008PyNNaCommonInterfaceforNeuronalNetworkSimulators}, Brian2 \cite{Stimberg2019Brian2anIntuitiveandEfficientNeuralSimulator}, MegaSim \cite{MegaSim}, SpiNNaker \cite{2020SpiNNakeraSpikingNeuralNetworkArchitecture} and Loihi \cite{Davies2018LoihiaNeuromorphicManycoreProcessorwithonChipLearning} where the SNN can be simulated or deployed. However, depending on the components used in the original ANN, some of the target platforms might not be available. 
During the conversion phase, Integrate-and-Fire (I\&F) neurons are used for a one-to-one substitution. These are then tuned so that their mean firing rate approximates the activation of the corresponding neuron in the original ANN. Neural networks are required to be pre-trained in their original framework as the toolbox does not provide any training utility. Instead, either through command line or through a simple GUI, it is possible to tune the conversion parameters and to perform inference on data.

\subsection{Lava}\label{sec:lava}
Lava \cite{Lava} is a relatively recent framework built by Intel's Neuromorphic Computing Lab (NCL). The framework is the result of an evolution from the Nx SDK software for Loihi chips, but aims to target other hardware platforms as well. Lava is composed of 4 main packages, namely Lava (core), Lava-DL, Lava Dynamic Neural Fields (DNF) and Lava Optimization. The current state of the platform includes the development of deep spiking neural networks trained with SLAYER \cite{Shrestha2018SLAYER}, and of SNNs converted from PyTorch-developed ANNs. Both SLAYER and the conversion system reside in the Lava-DL package. On-chip training through SLAYER is currently not available. Instead, models need to be trained off-chip and weights must be exported so that they can be used within the Lava core package. Within Lava-DL, a number of neuron models are defined, such as the LIF, Resonate-and-Fire (R\&F) \cite{Izhikevich2001ResonateandFireNeurons}, R\&F Izhikevich, Adaptive LIF \cite{Gerstner_2009}, Adaptive R\&F, and Sigma-Delta \cite{Cheung1993SigmaDeltaNeuralNetworks} modulation models. The core package currently supports LIF and Sigma-Delta modulation neurons. Recent developments in the framework have seen the implementation of on-chip learning functionalities through STDP and customized 3-factor learning rules.
Developing models in the core package requires understanding Lava's foundational concepts of Processes and ProcessModels. More specifically, Processes are the interfaces for every component of a system, have private memory and communicate through Ports. The ProcessModels are the implementations of these interfaces that can target different hardware.

\subsection{PyTorch-based Frameworks}
\subsubsection{Norse}\hfill

Norse \cite{Pehle2021NorseaDeepLearningLibraryforSpikingNeuralNetworks}
is a relatively recent PyTorch-based framework. It was developed with the aim of easing the construction of SNNs for ML solutions. This framework offers a wide range of neuron models, such as the LIF, LIF variants and extensions, and Izhikevich's model. It also provides a LSNN \cite{Bellec2018LSNN}, a spiking version of the LSTM (Long Short-Term Memory) \cite{hochreiter1997lstm}.
Norse has a functional programming style. Indeed, neurons are mainly implemented as functions and do not hold an internal state. Instead, the previous state of the neuron needs to be provided as an argument at each iteration.
The framework mainly allows for two types of learning: STDP \cite{masquelier2007SimplifiedSTDPRule} and SuperSpike \cite{zenke2018superspike}. Therefore, both local unsupervised learning and surrogate gradient learning are possible. Overall, Norse provides a good degree of flexibility and allows to leverage all of the features of PyTorch, such as execution on GPU.

\subsubsection{PySNN}\hfill

PySNN \cite{PySNN} is an other framework based on PyTorch aimed at the development of ML algorithms. Similarly to Nengo, connections between two neurons are modelled as separate objects that have properties and can affect the transmission of a signal. For instance, they can explicitly account for connection delays. 
Neuron models in PySNN embed the concept of spike trace, that can be used for learning purposes. Some available neuron models are the I\&F, LIF and ALIF. 
Concerning the learning rules, it is possible to use either STDP or MSTDPET (Modulated STDP with Eligibility Traces) \cite{Florian2007ReinforcementLearningthroughModulationofSpikeTimingDependentSynapticPlasticity}.
The framework also provides some useful utilities to load some NM datasets.
A downside of using PySNN is that the documentation is not complete, thus reading the source code itself is required to understand the working mechanics of the framework.

\subsubsection{SnnTorch}\hfill

An other framework basing its architecture on PyTorch is snnTorch \cite{eshraghian2021SnnTorch}. Connectivity between layers is enabled by leveraging PyTorch standard convolutional and fully connected layers. Spiking neurons are thought to be used as intermediate layers between these. Spiking neurons are modelled as classes that hold their own internal state. Available models include LIF-based models, second order LIF models, recurrent LIF models, and LSTM memory cells.
Learning in snnTorch takes place with BP Through Time (BPTT) using surrogate gradient functions to calculate the gradient of the spiking neurons. The framework also offers the possibility to use a Real-Time Recurrent Learning (RTRL) rule which applies weight updates at each time step, rather than at the end of a sequence of inputs. The output of the network can be interpreted using both a rate-based approach and a time-to-first-spike (TTFS) approach.
Finally, snnTorch provides access to the N-MNIST \cite{Orchard2015ConvertingStaticImageDatasetstoSpikingNeuromorphicDatasetsUsingSaccades}, DVS Gestures \cite{Amir2017ALowPowerFullyEventBasedGestureRecognitionSystem}, and the Spiking Heidelberg Digits \cite{Cramer2020TheHeidelbergSpikingDataSetsfortheSystematicEvaluationofSpikingNeuralNetworks} datasets, and includes useful network activity visualization tools.

\subsubsection{SpikingJelly}\label{sec:spikingjelly}\hfill

SpikingJelly \cite{Wei2020SpikingJelly} is a framework using PyTorch as a backend and adopting its coding style throughout. It provides implementations of I\&F, LIF, parametric LIF (pLIF), Quadratic I\&F (QIF), and Exponential I\&F neuron \cite{FourcaudTrocme2003HowSpikeGenerationMechanismsDeterminetheNeuronalResponsetoFluctuatingInputs} models. Firing of neurons in SpikingJelly is approximated by a surrogate function (such as the sigmoid) that allows differentiation and thus BP. The framework provides several utilities to read NM and non-NM datasets. Concerning the NM datasets, it is possible to both read them with a fixed integration time-window and with a fixed number of frames (every frame will have the same number of events). Among the available datasets, there are the CIFAR10-DVS \cite{Li2017CIFAR10DVSanEventStreamDatasetforObjectClassification} dataset, the DVS Gestures dataset, the N-Caltech101 \cite{Orchard2015ConvertingStaticImageDatasetstoSpikingNeuromorphicDatasetsUsingSaccades} dataset, and the N-MNIST dataset.
Finally, SpikingJelly also provides a functionality for ANN to SNN conversion from PyTorch.

\subsubsection{BindsNet}\hfill

BindsNet \cite{Hazan2018BindsNETaMachineLearningOrientedSpikingNeuralNetworksLibraryinPython} is a library for the development of biologically inspired SNNs and is based on PyTorch. However, despite having PyTorch as a backend, the coding style differs slightly. Execution is implemented by running the network for a certain amount of time on some input rather than explicitly looping through the dataset, making it more similar to Nengo in this regard. BindsNet supports several types of neuron models, among which I\&F, LIF, LIF with adaptive thresholds, Izhikevich's, and Spike Response Model (SRM)-based \cite{Gerstner_2009} models. Connections are modelled explicitly and link one node of the network with an other. Recurrent connections are also possible. The provided learning rules are biologically inspired and can be either two-factor (STDP or Hebbian) or three-factor (MSTDPET), hence no BP-based learning rule is proposed. Through subclassing, it is possible to customize neurons, input encoding and learning rules. 
The framework also provides utility tools to load datasets, such as the spoken MNIST, and DAVIS \cite{Brandli2014DAVIS} camera-based datasets. Finally, BindsNet includes a conversion system to convert neural networks developed in PyTorch into SNNs.

\subsubsection{SpykeTorch}\label{sec:spiketorch}\hfill

SpykeTorch is PyTorch-based library for building SNNs with at most one spike per neuron. This means that for each sequence of inputs, each neuron is allowed to fire only once. Because of this, tensor operations can be easily used to compute neuron activations. Because NM data includes the concept of time, what is normally treated as the batch dimension in PyTorch, it is interpreted as the time dimension in SpykeTorch. The framework is built to support STDP and Reward-modulated STDP (R-STDP) with a Winner Takes All (WTA) paradigm, and using convolutions as a connection scheme. The only available neuron model is the I\&F, which is provided as a function. Finally, the framework provides functionalities to encode non-NM input through difference of Gaussians and intensity to latency transforms, as well as some inhibition functions.

\section{SpykeTorch Spiking Neurons}\label{sec:spiking_neurons}

For the purpose of developing NM-ML algorithms based on STDP, SpykeTorch allows a high degree of customization and flexibility to the user. However, as mentioned in \ref{sec:spiketorch}, the framework originally provides a single spiking neuron model, the I\&F. This does not have a voltage leakage factor, which means that its internal state can only increase until it is reset. In order to augment the usage potential of SpykeTorch, we expand the library by implementing a new set of spiking neuron models, for a total of 8 new models, as show in Table \ref{tab:neurons}. By introducing more complex neuron models, the original workflow and implementation patterns adopted in the original framework cannot be easily utilized. Therefore, following are some details about the differences introduced to accommodate such neuron models in the library. We refer to the framework resulting from our changes as SpykeTorch-Extended.

\begin{figure}[h]
    \centering
    \includegraphics[width=\columnwidth]{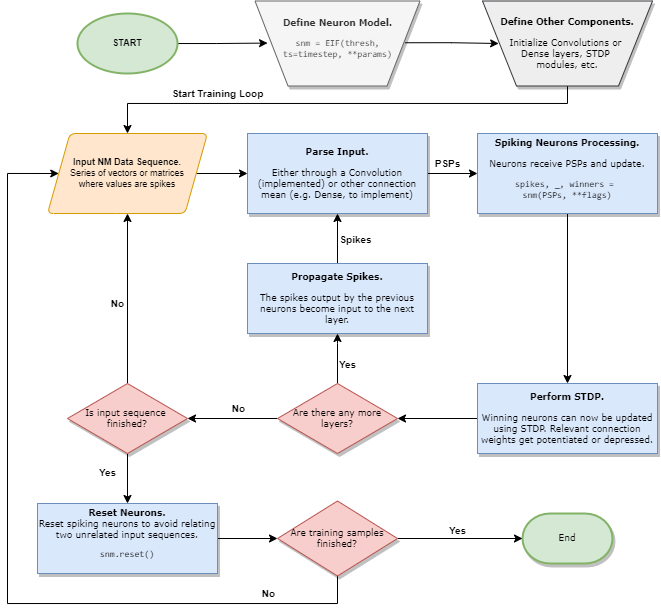}
    \caption{Example flowchart for SpykeTorch-Extended. After the definition of the components of the SNN, each data sample is required to be decomposed into its forming time steps before being processed by the SNN. After each single time step learning can take place, thus ensuring that learnt parameters will influence the result of the next iteration. It is also important to reset the state of the neuron layers after the last time step of each input sample to avoid unwanted effects.}
    \label{fig:flowchart}
\end{figure}

\subsection{Spiking Neurons Implementation Details}\label{sec:impl_details}
In our implementation of spiking neurons, we consider a subset from the phenomenological family of neuron models due to their computational efficiency \cite{Manna2022SimpleandComplexSpikingNeuronsPerspectivesandAnalysisinaSimpleSTDPScenario}. This is composed by the following:

\begin{multicols}{2}
    \begin{itemize}
        \item Leaky I\&F (LIF) \cite{Lapicque1907RecherchesQuantitativesSurLExcitationElectriqueDesNerfsTraiteeCommeUnePolarization}
        \item Exponential I\&F (EIF) \cite{FourcaudTrocme2003HowSpikeGenerationMechanismsDeterminetheNeuronalResponsetoFluctuatingInputs}
        \item Quadratic I\&F (QIF) \cite{Ermentrout1986ParabolicBurstinginanExcitableSystemCoupledwithaSlowOscillation}
        \item Adaptive Exponential I\&F (AdEx) \cite{Brette2005AdaptiveExponentialIntegrateandFireModelAsanEffectiveDescriptionofNeuronalActivity}
        \item Izhikevich's \cite{Izhikevich_2003}
        \item Heterogeneous Neurons.
    \end{itemize}
\end{multicols}
The LIF model is a single-variable neuron model and is the most widely used for the development of NM-ML systems \cite{Manna2022SimpleandComplexSpikingNeuronsPerspectivesandAnalysisinaSimpleSTDPScenario, VicenteSola2022KeystoAccurateFeatureExtractionUsingResidualSpikingNeuralNetworks, Mozafari2018FirstSpikeBasedVisualCategorizationUsingRewardModulatedSTDP, Hunsberger2015SpikingDeepNetworkswithLIFNeurons, Friedl2016HumanInspiredNeuroroboticSystemforClassifyingSurfaceTexturesbyTouch}; the EIF and QIF models are other single-variable models  that include different types of complexities in their equation, and are also the base for more complex models, the AdEx and Izhikevich's respectively; the AdEx and Izhikevich's models are two-variable neuron models that have also been widely studied and employed in the literature \cite{Schemmel2020AcceleratedAnalogNeuromorphicComputing, Barton2018TheApplicationPerspectiveofIzhikevichSpikingNeuralModeltheInitialExperimentalStudy, Chaturvedi2014ReviewofHandwrittenPatternRecognitionofDigitsandSpecialCharactersUsingFeedForwardNeuralNetworkandIzhikevichNeuralModel}.

Due to the greater complexity of the newly introduced neurons, we deviate from the original implementation and adopt an object-oriented approach for the neurons. This allows them to retain an internal state and other properties. Nevertheless, to maintain compatibility, neuron objects are callable and share the same output format as in the original implementation. Furthermore, we do not restrict neurons to firing only once per input sequence. This only depends on the choice of parameters for a given neuron, such as the refractory period. Another difference with the previous implementation is that the neurons are expected to receive events one time-step at a time. While this introduces a overhead on the computational time, it allows to simulate real-time processing, and also ensures the decay of the membrane potential and that weight updates due to STDP affect every subsequent moment in time, thus making the system more realistic. 
A neuron layer in SpykeTorch-Extended is characterized by at least the following parameters: a time constant \code{tau\_rc}, a time-step size \code{ts}, a capacitance \code{C}, a \code{threshold}, a \code{resting\_potential}, and by the number of \code{refractory\_timesteps}. These are all the required parameters for a LIF neuron; however, more complex neuron models will require more parameters. A layer of neurons in this system can be better depicted as a set of neuronal populations. The number and size of the population reflect that of the input that is processed by the layer. A single population is thus intended as being the group of neurons that correspond to one of the feature maps produced by a convolutional layer. In the case of a dense layer, each population is composed of one neuron only.

As a result of the changes above, the standard workflow in SpykeTorch-Extended requires some adjustments with respect to the original version. In Figure \ref{fig:flowchart}, we report an example flowchart of how a pipeline using the new neuron models could look like. As the flowchart highlights, each input is expected to be unravelled into all the time steps it is composed of and, for each time step, all the events that took place in such a time span are to be fed forward to the SNN. 




\begin{table}[]
\caption{Summary of newly added spiking neurons to SpykeTorch. All the neurons share a base set of parameters with the LIF, but they may require more depending on the neuron type, which are briefly reported in the short description.}
\label{tab:neurons}
\begin{tabularx}{\textwidth}{cX}
\toprule
\textbf{Neurons}   & \multicolumn{1}{c}{\textbf{Short Description}}                                                                                                                                                                   \\ \midrule
\textbf{LIF} \cite{Lapicque1907RecherchesQuantitativesSurLExcitationElectriqueDesNerfsTraiteeCommeUnePolarization}       & Uses the integral solution to the differential equation in \cite{Gerstner_2009}.                                                                                                                                 \\ \midrule
\textbf{EIF} \cite{FourcaudTrocme2003HowSpikeGenerationMechanismsDeterminetheNeuronalResponsetoFluctuatingInputs}       & Single-variable model featuring an exponential dependency from the membrane potential. Has parameters \code{delta\_t} for the sharpness of the curve, and \code{theta\_rh} as a cut-off threshold for the upswing of the curve \cite{Gerstner_2009}.                                                      \\ \midrule
\textbf{QIF} \cite{Ermentrout1986ParabolicBurstinginanExcitableSystemCoupledwithaSlowOscillation}       & Single-variable model featuring an quadratic dependency from the membrane potential. Has parameters \code{a} for the steepness of the quadratic curve, and \code{u\_c} as the negative to positive updates crossing point of the membrane potential \cite{Gerstner_2009}.                              \\ \midrule
\textbf{AdEx} \cite{Brette2005AdaptiveExponentialIntegrateandFireModelAsanEffectiveDescriptionofNeuronalActivity}      & Two-variables model similar to the EIF, but with an adaptation variable. It adds parameters \code{a} and \code{b}, respectively for adaptation-potential coupling and adaptation increase upon spike emission.                                                                 \\ \midrule
\textbf{IZ} \cite{Izhikevich_2003}        & Two-variables model similar to the QIF, but with an adaptation variable. It adds parameters \code{a} for the time scale of the adaptation variable, \code{b} for the subthreshold sensitivity of the adaptation, and \code{d} for the adaptation increase upon spike emission. \\ \midrule
\textbf{H-Neurons} & Heterogeneous versions of LIF, EIF, and QIF neurons with uniformly distributed \code{tau\_rc} parameter.                                                                                                          \\ \bottomrule
\end{tabularx}
\vspace{-2.5pt}  
\end{table}

\subsection{Heterogeneous Neuron Classes}
The implemented neuron classes create a layer of spiking neurons that share the same hyper-parameters. We refer to this as being a homogeneous layer of neurons because they all react in the same way to the same sequence of inputs. However, it might be useful to have neurons reacting differently to one input, since this could mean being able to learn different kinds of temporal patterns within the same layer. Because of this, we further implement heterogeneous neuron classes for the \code{LIF}, \code{EIF}, and \code{QIF} classes. Specifically, they provide a set of $\tau_{rc}$ values that are uniformly distributed within a range specified by the user through the parameter \code{tau\_range}. We limited the current implementation to a uniform distribution for simplicity, and limit the heterogeneity to the $tau_{rc}$ parameter since this directly influences the time scale to which the neuron is sensitive. Nevertheless, future developments will consider other types of distributions and parameters.

\section{Conclusions}\label{sec:conclusions}
In this work we have presented a review of 9 Python frameworks for the development of spiking neural networks oriented towards data science applications. We have seen that several of them use PyTorch as a base to leverage the GPU acceleration, to exploit the existing functionalities it offers, and to ease the transition for users that come from a conventional DL background. Nevertheless, they all differ slightly in their implementations and in the SNN development tools they offer. Other frameworks like Nengo and Lava do not have such a base, but provide conversion methods to increase usability. This review also highlights how, despite restricting our field of view to data science-oriented libraries, there is a wide variety of frameworks. This is possibly due to growing interest that SNNs have lately received, however this also reflects the lack of an established and widespread framework like in the case of PyTorch or TF/Keras for conventional DL. 
Finally, we report our extension to a specific framework, SpykeTorch, that includes several new spiking neurons to use for simulations. Our additions require a modification of the original workflow, but enable real-time processing simulation with STDP. By doing this, we hope to promote and speed up future research in this direction, as well as to contribute to the development of richer software frameworks.

\newcommand{\newblock}{}
\bibliographystyle{unsrt}
\bibliography{./frameworks.bib}

\end{document}